\documentclass[11pt,twoside]{article}
\usepackage{asp2010}

\resetcounters

\begin{document}

\title{High-Resolution Sub/millimeter Observations of Mergers and Luminous Galaxy Nuclei}
\author{Kazushi~Sakamoto
\affil{Academia Sinica Institute of Astronomy and Astrophysics, P.O. Box 23-141, Taipei 10617, Taiwan}}

\begin{center}
\fontsize{10pt}{10.5pt} \selectfont
Presented at the conference titled \\
``GALAXY MERGERS IN AN EVOLVING UNIVERSE'' (2011. Oct., Hualien). \\
%To appear in the ASP Conference Series.
ASP Conference Series, in press.
\end{center}

%%%%%%%%%%
\begin{abstract}
I present recent high-resolution submillimeter and millimeter observations of molecular gas 
and dust in some mergers, luminous galaxy nuclei, and possible mergers. 
Such observations tell us the behavior and properties of interstellar medium in merger nuclei.
For example, the gas sometimes makes a mini disk around the remnant nucleus, 
feeds starburst and/or a massive black hole there, hides such a power source(s) by enveloping it, 
and is blown out by the embedded power source. 
Even when the power source is completely enveloped and hidden
we can still constrain its physical parameters and nature from
high-resolution (sub)millimeter observations. 
The observables include  
gas motion such as rotation (hence dynamical mass) and inflow/outflow, 
luminosity and luminosity density of the embedded nucleus, 
and
mass, temperature, density, chemical composition, and (sometimes unusual) excitation conditions
of gas.
\end{abstract}

%%%%%%%%%%%%%%%%%%%%%%%%%%%%%%%%%%%%%%%%%%%%%%%%%%%%
\section{Introduction}
Among various ways to observationally study mergers,  high-resolution
observations at submillimeter and millimeter wavelengths can shed unique light
on the inner working and evolution of mergers.
This is partly because the above-mentioned observations provide detailed spatial and kinematical
information of molecular gas that is the dominant interstellar medium in the inner regions of mergers.
It is also because star formation and active nuclei, the main energy-generating mechanisms in mergers,  
rely on the interstellar molecular gas for raw material or fuel. 
We can examine this important ISM at the very region where star formation
is most active or a massive black hole is being fueled, owing to recent progress in the 
high-resolution observing capabilities at the (sub)millimeter wavelengths.

I show below the power of high-resolution (sub)millimeter observations in three areas
related to mergers.
First, compact luminous nuclei are often, though not exclusively, seen in major mergers.
Submillimeter imaging of these nuclei at subarcsecond resolution can reveal 
hidden properties of young AGNs or starbursts in these nuclei.
Examples of such nuclei in Arp 220 and NGC 4418 are presented.
Second, outflow of molecular gas from mergers can be found thorough 
high-resolution observations of (sub)millimeter molecular lines. 
Such outflows can be driven by extreme starburst or AGN in mergers and likely affect the evolution of these activities.
Examples of such outflows are presented for Arp 220 and NGC 3256.
Third, subtle effects of a minor galaxy merger may be detectable with high resolution observations of
molecular gas. 
Observations of NGC 4418 and M83 are addressed as possible cases where minor
merger may have played a role.

%%%%%%%%%%%%%%%%%%%%%%%%%%%%%%%%%%%%%%%%%%%%%%%%%%%%
\section{Compact Luminous Nuclei}
Enhanced luminosity is probably next to morphological changes among the most visible effects 
merging can have on galaxies.  
Arp 220 is the nearest ultraluminous infrared galaxy 
and is an advanced merger with two nuclei separated by only 300 pc on the sky.
NGC 4418 is an infrared luminous disk galaxy though less luminous than Arp 220 
($L_{\rm 8-1000\, \micron} = 10^{11.1} L_{\odot}$ while Arp 220 has $10^{12.2} L_{\odot}$, \citealt{Sanders03}). 
Both have heavily absorbed nuclei most clearly indicated in their mid-IR spectra where both have
a deep absorption feature around 10 \micron\ due to silicate dust \citep{Spoon07}.
Almost complete absorption of the 10 \micron\ continuum implies that the nuclei generating most of the 
10 \micron\ luminosity are almost completely covered by a thick layer of dust and are compact enough to be fully covered. 
Therefore it has been suggested for a long time that they host an AGN or a very compact young starburst
\citep{Roche86,Smith89}.
The following examples show how observations of gas and dust at high resolution can improve our 
understanding of these nuclei and their relation to mergers.

%%%%%%%%%%
\subsection*{SMA Submillimeter Imaging of Arp 220}
\begin{figure}[t]
\plotone[width=5.0cm]{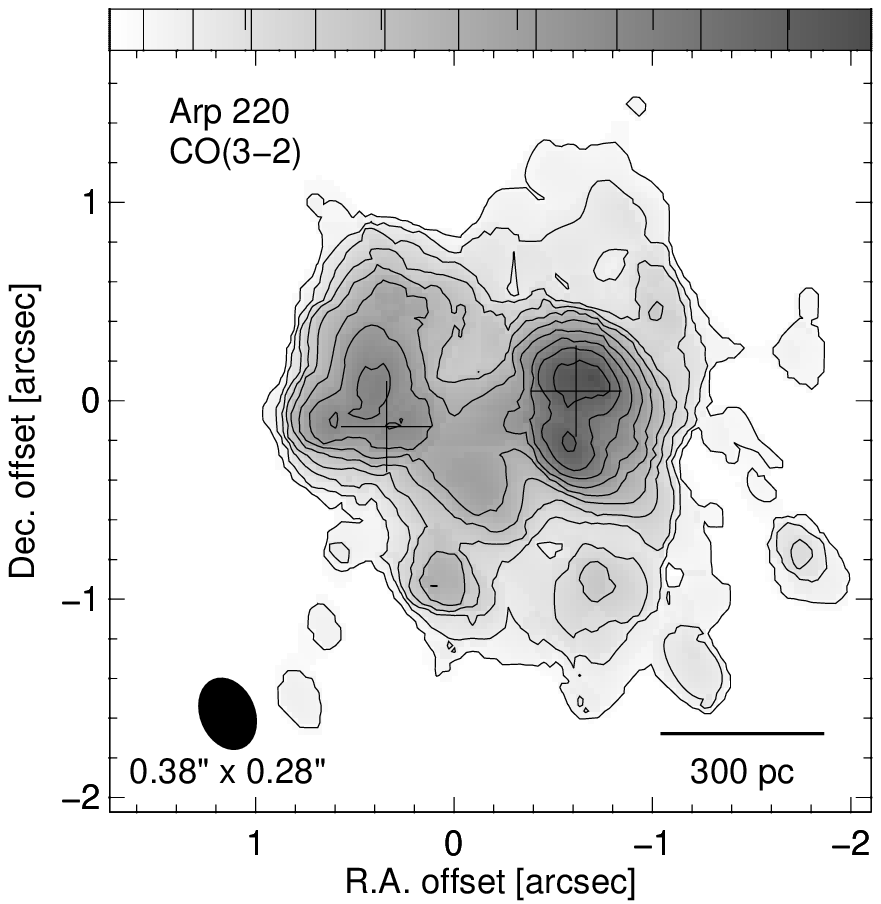} 
\plotone[width=5.0cm]{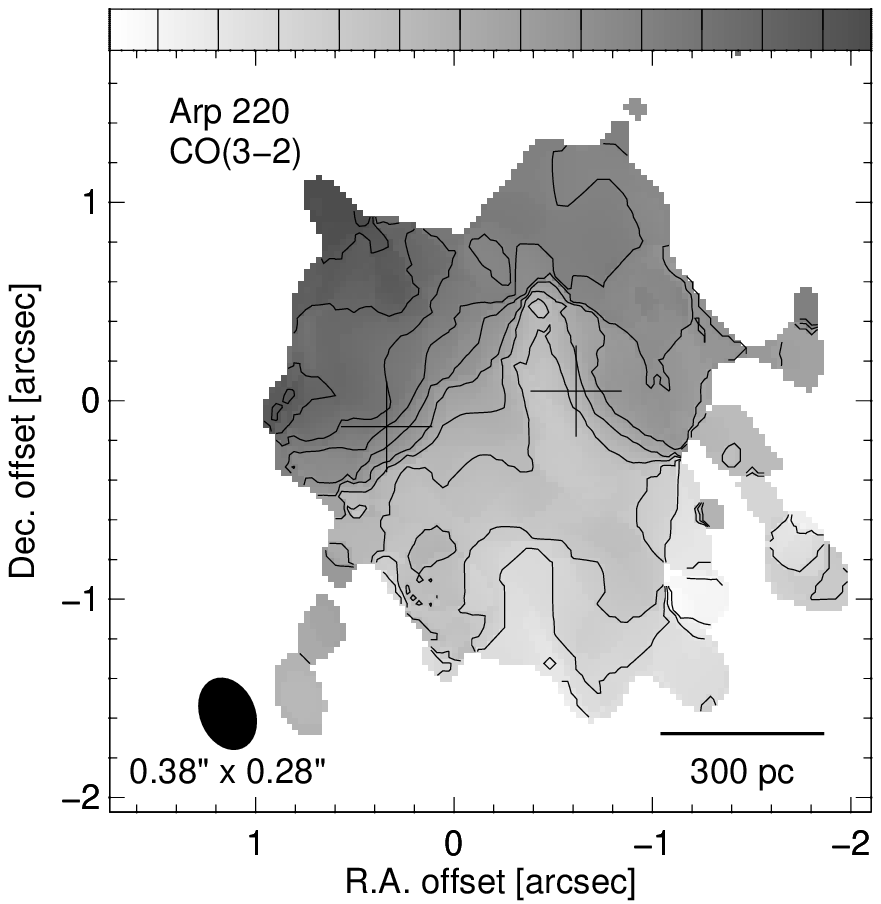}
\plotone[width=5.0cm]{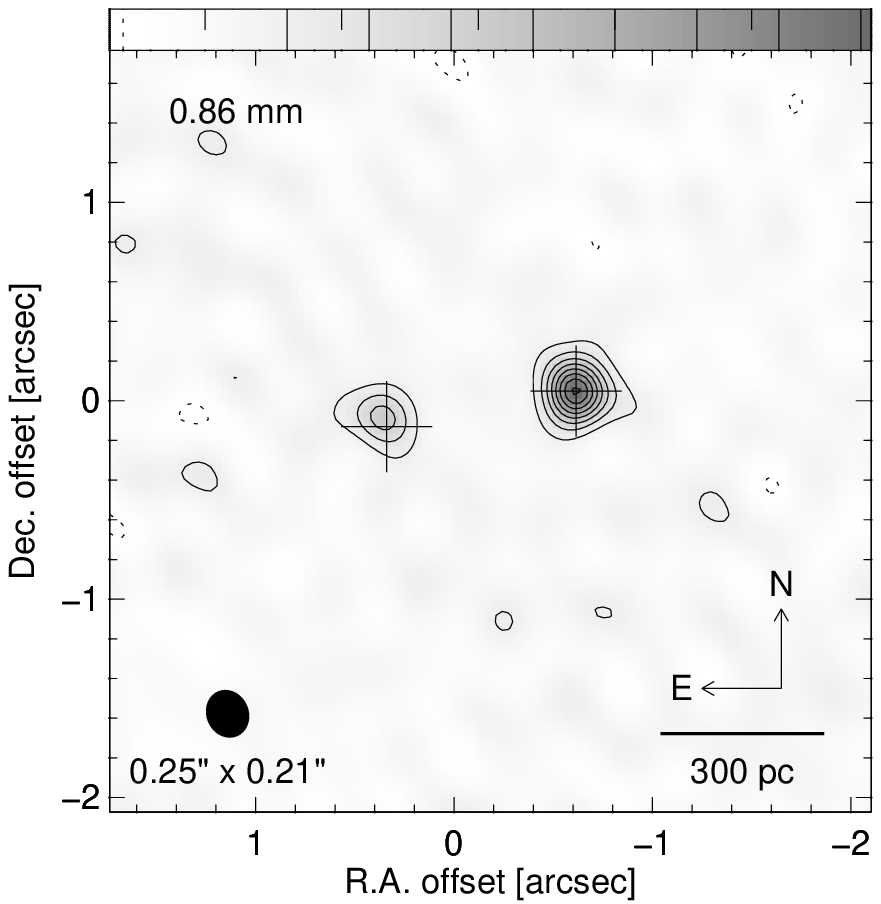}  
\plotone[width=4.7cm]{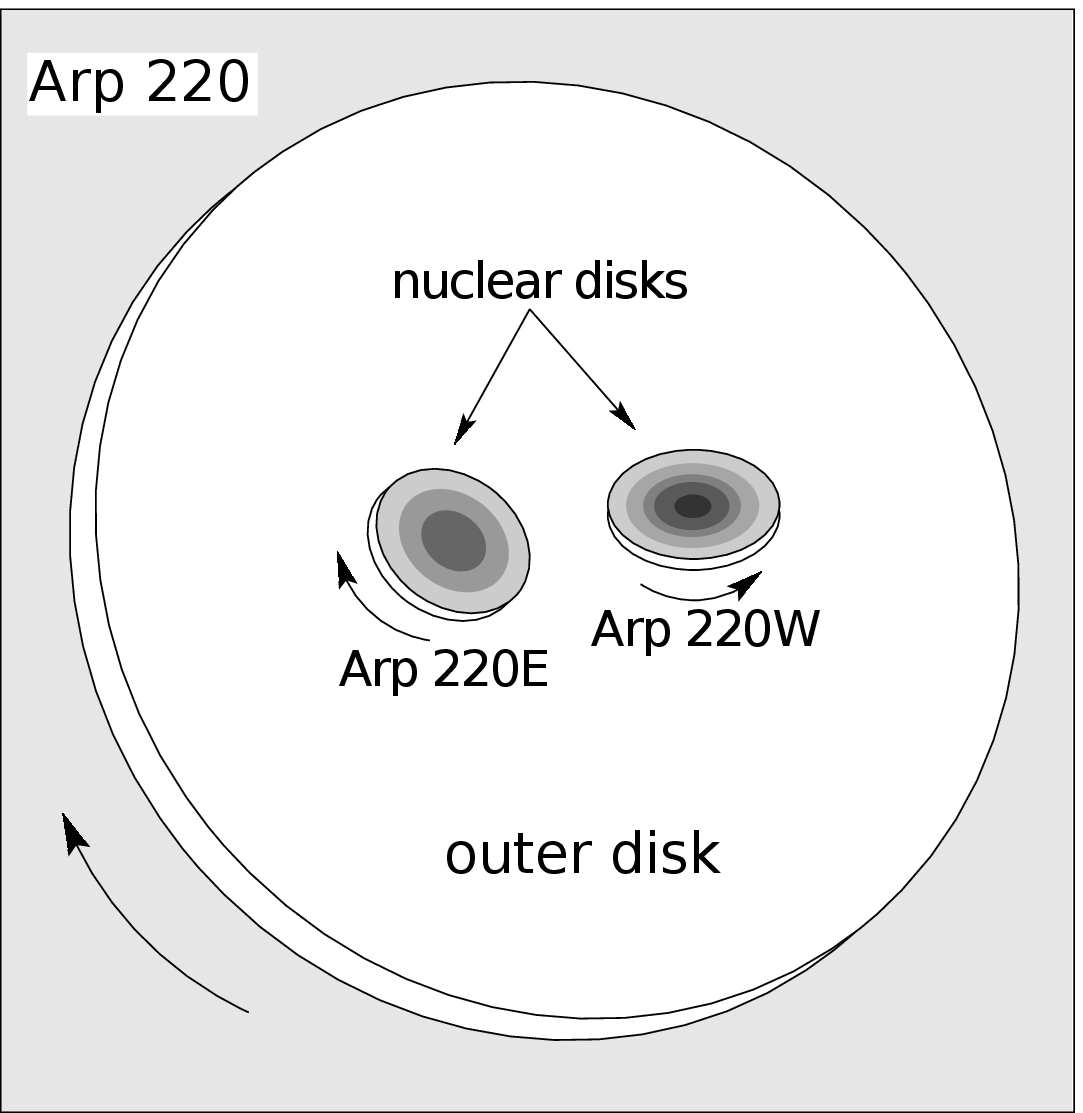} 
\caption{Molecular gas and dust emission in the center of Arp 220 merger.
(top-left) CO(3--2) integrated intensity. (top-right) CO(3--2) velocity field. The velocity contours are in 50 km s$^{-1}$ steps.
(bottom-left) 860 $\mu$m continuum emission from dust.
These 350 GHz data are from \citet{Sakamoto08}.
(bottom-right) An illustration of the Arp 220 system. 
Each of the two nuclei has its own nuclear disk and the binary nucleus is surrounded by 
a larger molecular disk \citep{Sakamoto99}. 
\label{fig.arp220maps} }
\end{figure}

Fig. \ref{fig.arp220maps} shows the first subarcsecond resolution submillimeter images of Arp 220
\citep{Sakamoto08}, obtained with the Submillimeter Array (SMA\footnote{The Submillimeter Array is a joint
project between the Smithsonian Astrophysical Observatory and the
Academia Sinica Institute of Astronomy and Astrophysics, and is
funded by the Smithsonian Institution and the Academia Sinica.
}). 
The CO emission shows concentrations of molecular gas at the two nuclei and an envelope of 
molecular gas around the binary nucleus. 
Each nucleus has its own mini gas disk showing rotational velocity gradient.
The rotational axes are misaligned with each other.  
The continuum emission from dust is much more compact than the CO line emission. 
The brighter nucleus, Arp 220W, has a deconvolved continuum size of a couple 
of tenths of arcsecond or 50--80 pc.
The continuum emission is very bright and has a peak deconvolved
brightness temperature around 100 K. 
This is a lower limit to dust temperature because of the opacity of the dust continuum.
From these we could set a lower limit to 
the luminosity of the nucleus at (2--3)$\times 10^{11} L_{\odot}$ 
using the Stefan-Boltzmann formula.
This lower limit is quite significant.
For example, the luminosity surface density of the nucleus is $\geq$$4\times 10^7 M_{\odot}$ ${\rm pc}^{-2}$.
The luminosity to mass ratio of the nucleus, $\geq$400 $L_{\odot}/M_{\odot}$, is also high,
where the dynamical mass of the nucleus was estimated to be 
$M_{\rm dyn}(r\leq 40\, {\rm pc})\approx 6\times10^{8}\;M_{\odot}$
from the CO velocity information.

In all, Arp 220 has a textbook case of merger-induced galaxy activities in the sense 
that the major merger funneled a huge amount of gas to the galaxy nuclei and triggered a starburst
and plausibly a quasar-like AGN activity.
Rather than hasting to conclude whether the nucleus hosts an energetically
dominant AGN or not, although it seems likely, I stress that 
an important lesson here is that we can measure many parameters of
compact, luminous, and heavily absorbed merger nuclei through high-resolution submillimeter observations,
which ALMA will further improve soon.

%%%%% Fig. Arp 220
%%%%%

%%%%%%%%%%
\subsection*{SMA Submillimeter Imaging of NGC 4418}
The case of NGC 4418 may be a caution to us that things are not always as simple as in Arp 220.
This galaxy shares a deep silicate absorption with Arp 220 
and, when observed with the SMA in the same way,  
the nucleus looks similar to those in Arp 220 although NGC 4418
does {\it not} appear to be a recent major merger.

\begin{figure}[t]
\plotone[width=4.5cm]{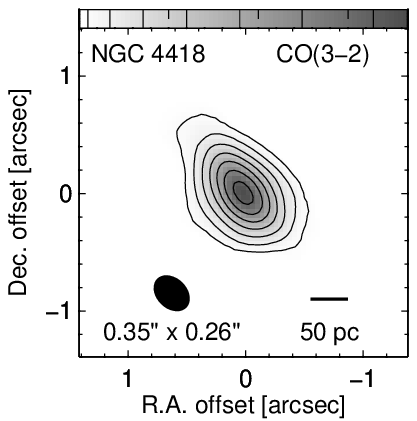} 
\plotone[width=4.5cm]{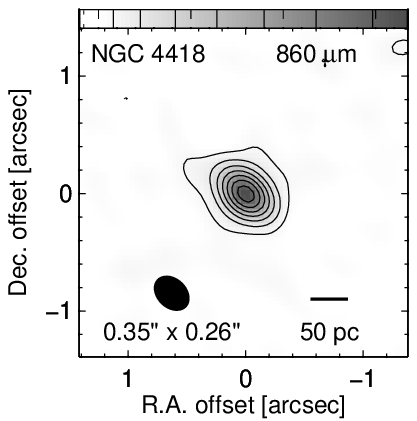}  
\caption{Molecular gas and dust emission at the nucleus of NGC 4418 (Sakamoto et al. in prep.).
\label{fig.ngc4418maps}}
\end{figure}

SMA images of NGC 4418 in Fig. \ref{fig.ngc4418maps} revealed
a strong nuclear concentration of molecular gas.  
The gas is unusually warm, having a peak CO brightness temperature in excess of 80 K. 
In addition we detected at the nucleus, for the first time in an external galaxy, rotational molecular lines
from vibrationally excited HCN, the energy level of which is more than 1000 K above the ground \citep{Sakamoto10}. 
As in Arp 220, the submillimeter continuum emission from the nucleus is compact.
The continuum core has a size of about 20 pc and a brightness temperature of about 100 K or higher.
Depending on the dust opacity of the nucleus, the core can easily be the overwhelmingly
dominant luminosity source of the entire galaxy.

In all, the luminous nucleus of NGC 4418 is similar to the Arp 220 nuclei or at least to the western
nucleus of Arp 220. 
But NGC 4418 is not a major merger unlike Arp 220.
Thus the high concentration of gas to this nucleus and the triggering of possible AGN or compact starburst there
cannot be attributed to gas funneling due to a major merger.
The cause of this gas structure is a remaining problem in order for us to put active mergers like Arp 220 
in the context of more general galaxy population.

%%%%% Fig. N4418

%%%%%%%%%%%%%%%%%%%%%%%%%%%%%%%%%%%%%%%%%%%%%%%%%%%%
\section{Molecular Winds and Outflows from Mergers}
Winds and outflows of molecular gas from mergers can be also found and examined  
using (sub)millimeter molecular lines. 
These outflows remove the gas concentrated to the merger nuclei and are potentially important in the merger evolution 
because they contribute to the emergence of the obscured nuclear activities and 
may eventually quench the activities by removing their fuel.

%%%%%%%%%%
\subsection*{Molecular P-Cygni Profiles toward Arp 220 Nuclei}
\begin{figure}[t]
\plotone[width=6.0cm]{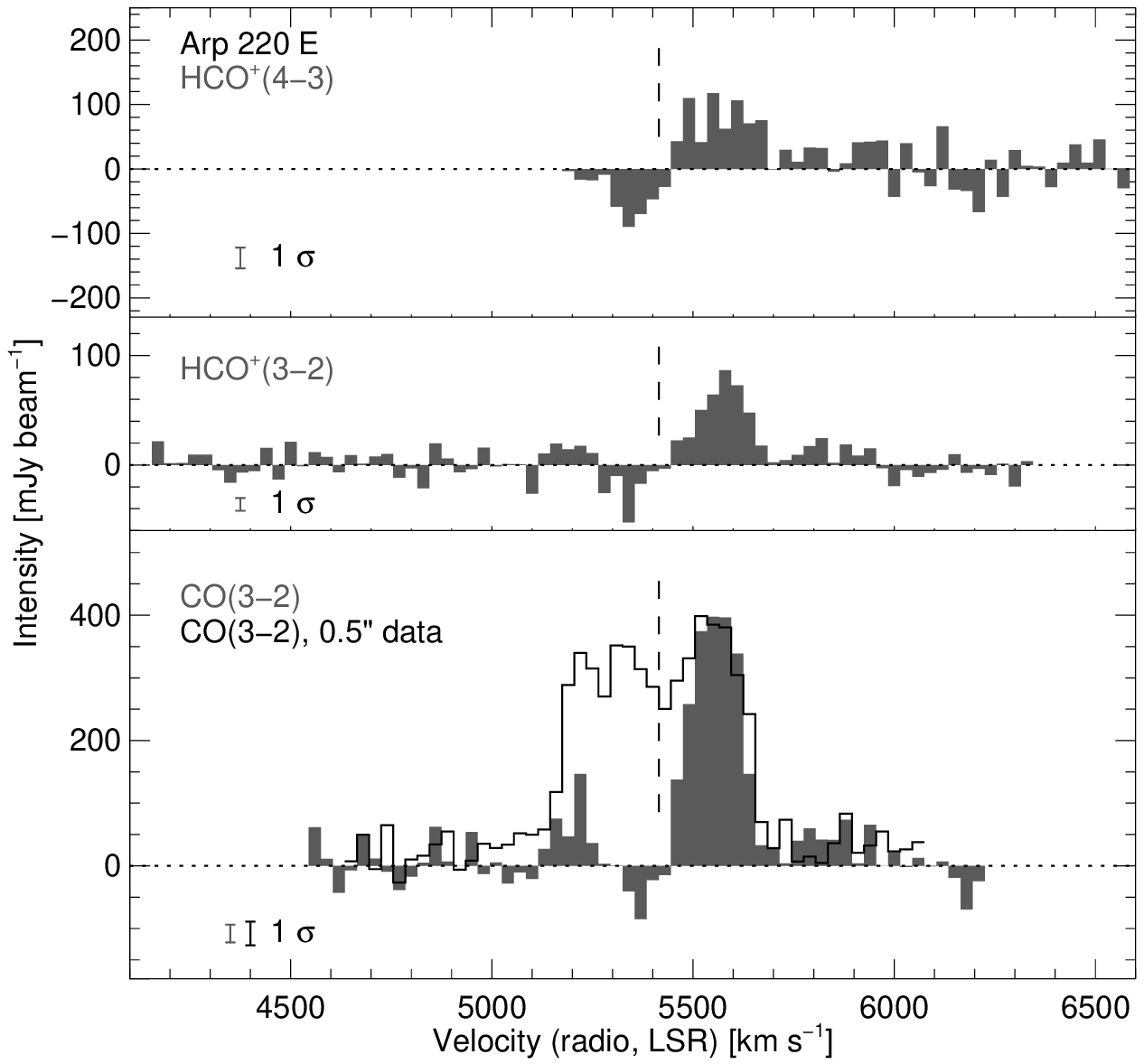}  
\plotone[width=6.0cm]{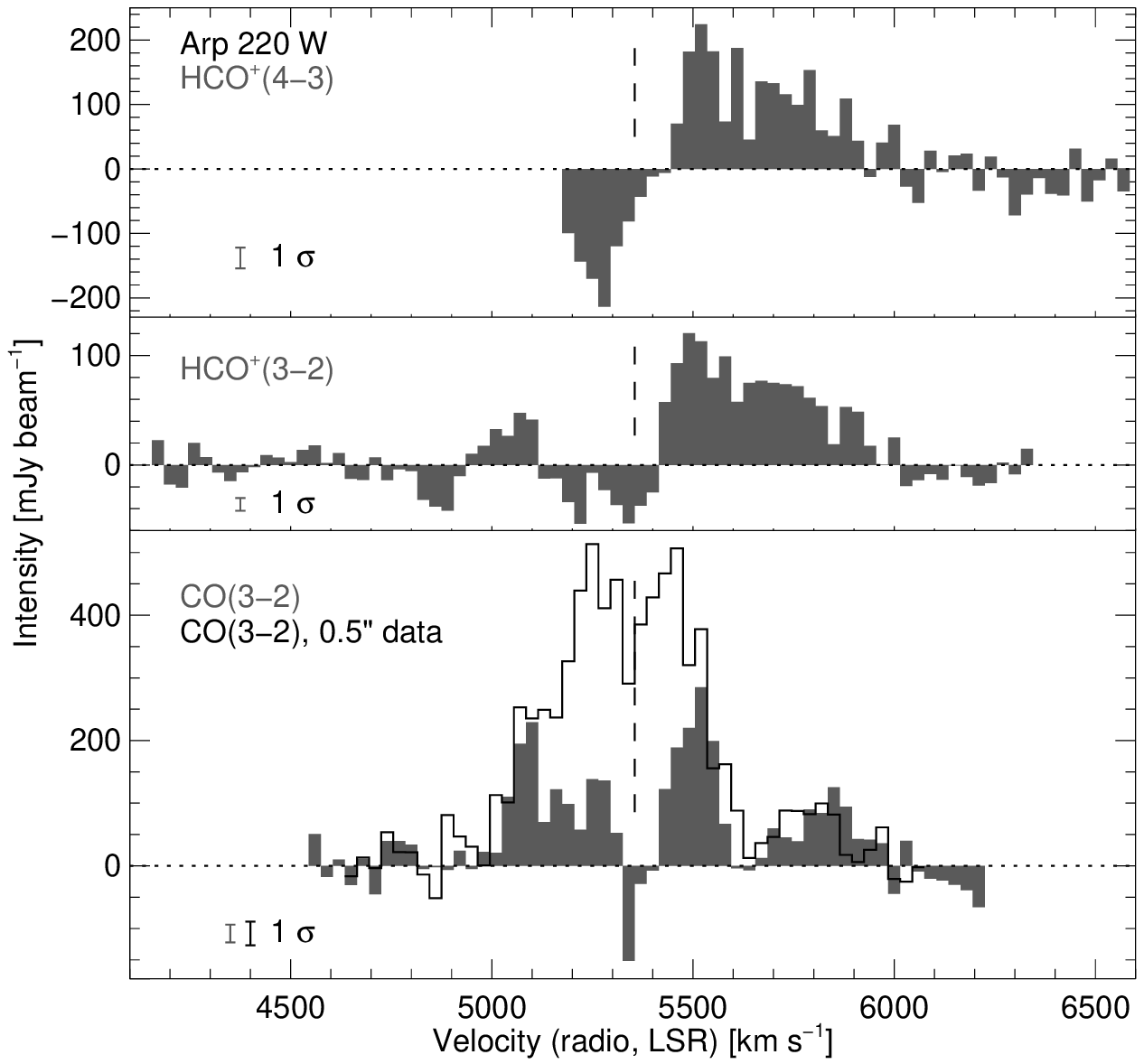}  
\caption{Profiles of (sub)millimeter molecular lines observed 
toward the nuclei of Arp 220; (left) eastern nucleus, (right) western nucleus.
Profiles shown in gray are observed with $\sim$$0.3''$ resolution.
The vertical dashed lines are at the systemic velocities of individual nuclei determined
from lower resolution ($0.5''$) CO(3--2) spectra.
Both nuclei show P-Cygni profiles indicative of molecular outflows \citep{Sakamoto09}.
\label{fig.a220PCyg}}
\end{figure}

Fig. \ref{fig.a220PCyg} shows the P-Cygni line profiles that we found toward 
the nuclei of Arp 220 from 0.3\arcsec\ resolution observations \citep{Sakamoto09}.
The molecular line profiles shown in gray have absorption at and on the low-velocity side of 
the systemic velocity and emission mostly on the high-velocity side.
This P-Cygni type asymmetry is more prominent in the HCO$^{+}$ profiles than in CO, and is unnoticeable
in the CO line profile obtained at a lower 0.5\arcsec\ resolution. (This shows importance of high resolution.)
The systemic velocities of individual nuclei were determined as the centroid velocities from
these lower resolution CO data and are shown as vertical dashed lines.
The blueshifted absorption is against the the bright continuum nuclei discussed above. 
Thus the absorbing gas must be in front of the nuclei, i.e., between us and each nucleus.
The blueshift of the absorption means that the absorbing gas is moving toward us.
Thus the simplest interpretation of the P-Cygni profiles is that the gas around each nucleus 
is moving away from the continuum core, or flowing outwards, in both nuclei. 
This is indeed the normal interpretation of P-Cygni line profiles.
Our observations thus suggested  that each of the two nuclei has its own outflow form its nuclear disk.
The molecular outflow rate was estimated to be on the order of 100 $M_{\odot}\; {\rm yr}^{-1}$ although this
strongly depends on the poorly constrained outflow geometry and  gas properties.

%%% Fig. P-Cyg profiles of Arp 220

%%%%%%%%%%
\subsection*{High-Velocity CO from NGC 3256 Merger}
\begin{figure}[t]
\plotone[width=6cm]{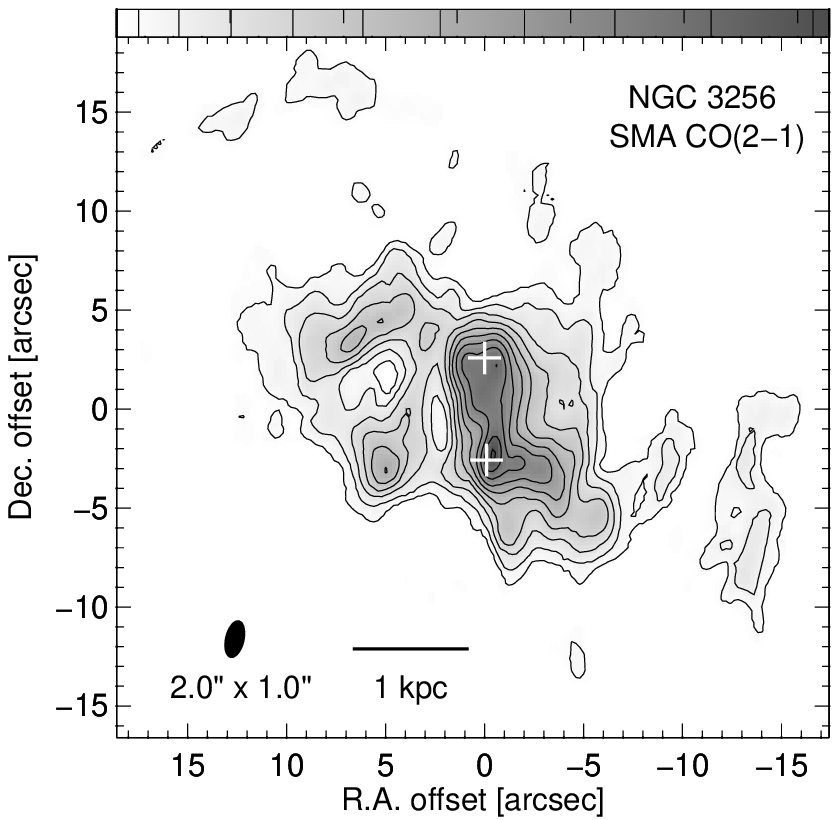} 
\plotone[width=7cm]{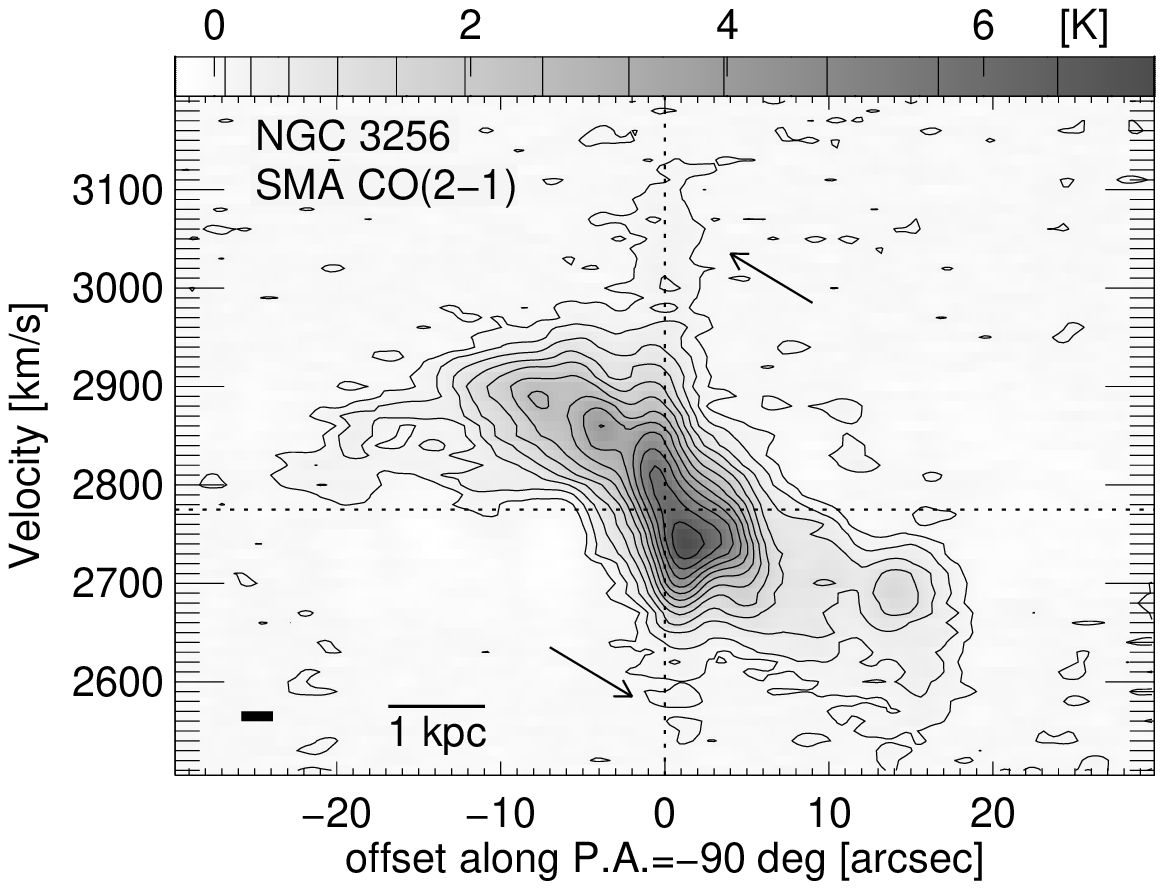}  
\plotone[width=7cm]{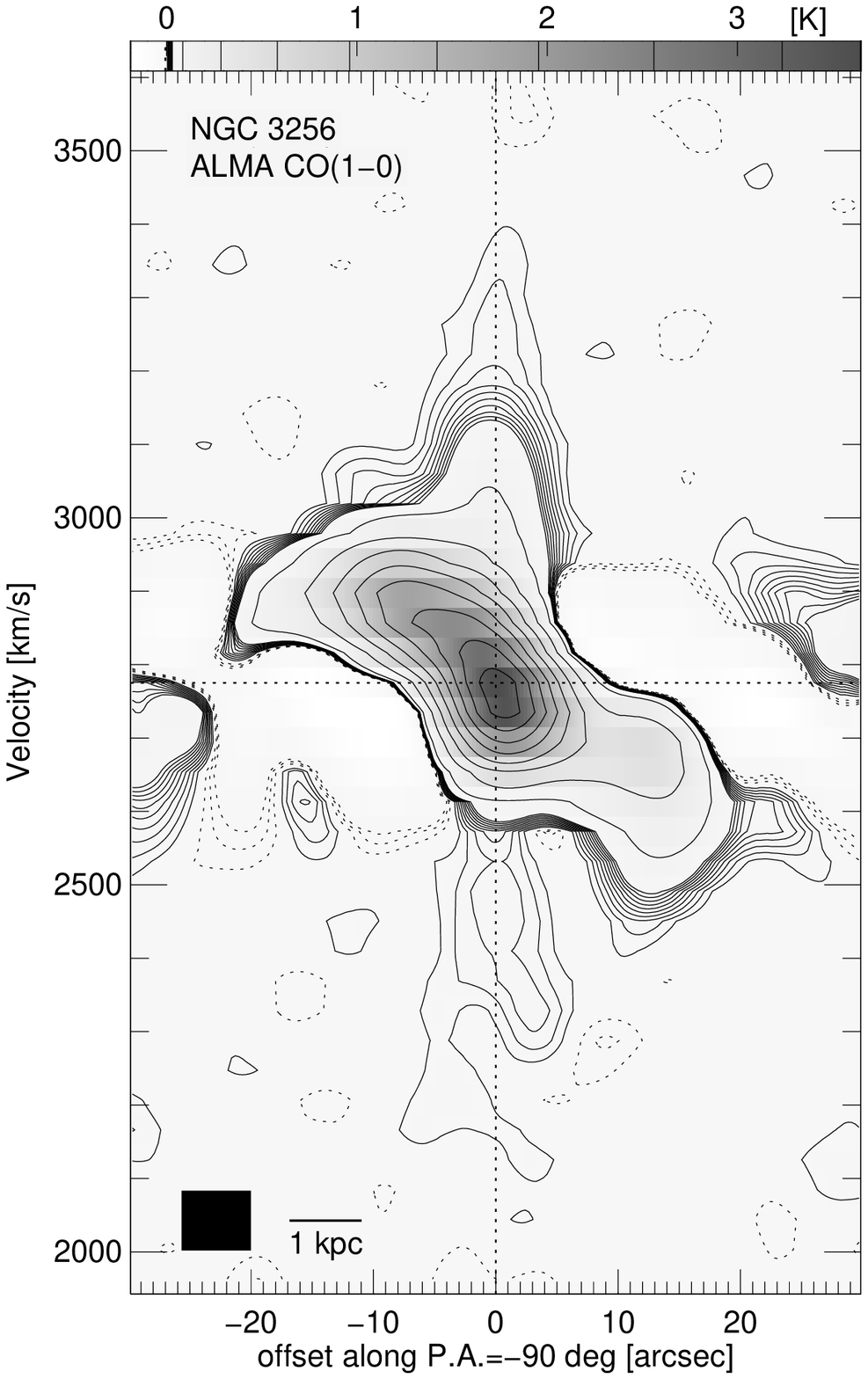}  
\caption{Molecular gas in the merger NGC 3256. 
(top-left) CO(2--1) integrated intensity. 
There are gas concentrations toward the two merger nuclei marked with white plus signs 
and spiral features around them.
(top-right) CO(2--1) position-velocity diagram showing the high-velocity gas in the nuclear region.
This is equivalent to a slit spectrum taken with a 6$''$-wide slit  covering both nuclei
and placed in the east-west direction.
The CO(2--1) data are from \citet{Sakamoto06}.
(bottom) CO(1--0) position-velocity diagram. 
It reveals that the high-velocity CO is more than 1000 km s$^{-1}$ wide. 
The CO(1--0) data are from the ALMA archive.
\label{fig.N3256}}
\end{figure}

One can also infer molecular outflow from galaxies only using molecular line emission.
The upper panels of Fig. \ref{fig.N3256} show SMA observations of a merger NGC 3256
\citep{Sakamoto06}. 
NGC 3256 is the most luminous galaxy within $z$=0.01 and is a merger having two close nuclei,
similar to Arp 220.
The SMA observations revealed two concentrations of molecular gas
at the locations of the two merger nuclei. 
The gas configuration is reminiscent of that in Arp 220.
In addition, a high-velocity CO emission was discovered at the center of the merger.
It is at the center of the CO(2--1) position-velocity diagram in Fig. \ref{fig.N3256} 
and shows the total line width of about 600 km s$^{-1}$.
The broad CO line was interpreted as a molecular outflow from the center of the merger.
The outflow rate was estimated to be on the order of 10 $M_{\odot}\; {\rm yr}^{-1}$.
Because this is comparable to the star formation rate in NGC 3256, the molecular outflow
is as significant as star formation in the gas consumption budget of the merger.

The broad CO line wings are confirmed and found to be broader 
in ALMA science verification data.
The lower panel of Fig. \ref{fig.N3256} shows a position-velocity diagram made
from the ALMA CO(1--0) data. 
The ALMA data are an order of magnitude more sensitive in brightness temperature than the SMA data
presented above.  
The high-sensitivity ALMA data not only confirm the presence of the high-velocity CO 
but also reveal that the line is more than twice wider than we saw with the SMA.
This means that the molecular outflow has a larger magnitude than we previously thought.
Thus the effect of the outflow may well exceed that of star formation in the gas spending budget
of the merger.
These observations are followed up with our ALMA Cycle 0 project that will
better constrain the full extent in space and velocity of the high-velocity gas
and the structure and the driving source(s) of the presumed outflow.
The two nuclei may be individually driving an outflow because the highest velocity gas
in the ALMA data is associated with the individual nuclei.
There may be two outflows at least at the source as inferred in Arp 220.

There have been many recent detections of galactic molecular outflows thorough
molecular P-Cygni profiles \citep[e.g.,][]{Fischer10} and
high-velocity CO wings \citep[e.g.,][]{Feruglio10,Chung11}.
\citet{Chung11} detected $\sim$1000 km s$^{-1}$ CO wings in the stacked spectrum of
14 local star-forming ULIRGs.
The population of galaxies and mergers with significant molecular outflows is rapidly growing.
The effect of molecular outflow may well be much more significant for the galaxy/merger evolution 
than we knew a few years ago.

%%%%% Fig. line-wings of NGC 3256

%\clearpage   % may need this to place figures at intended positions

%%%%%%%%%%%%%%%%%%%%%%%%%%%%%%%%%%%%%%%%%%%%%%%%%%%%
\section{Minor Mergers}
The effect of individual minor merger is more subtle and difficult to observe than that of a major merger even
though the integral effect of minor mergers may be significant in galaxy evolution.
Still we may be able to detect and study such subtle effects with
high resolution observations of molecular gas sensitive to small scale features.
For example, one can hypothesize and try to verify that the compact nuclear gas concentration in NGC 4418,
a galaxy without severe disturbance, may be due to such a minor merger. 
I here show one more case of molecular gas observations where a minor merger may have played a role.

\begin{figure}[t]
\plotone[height=6cm]{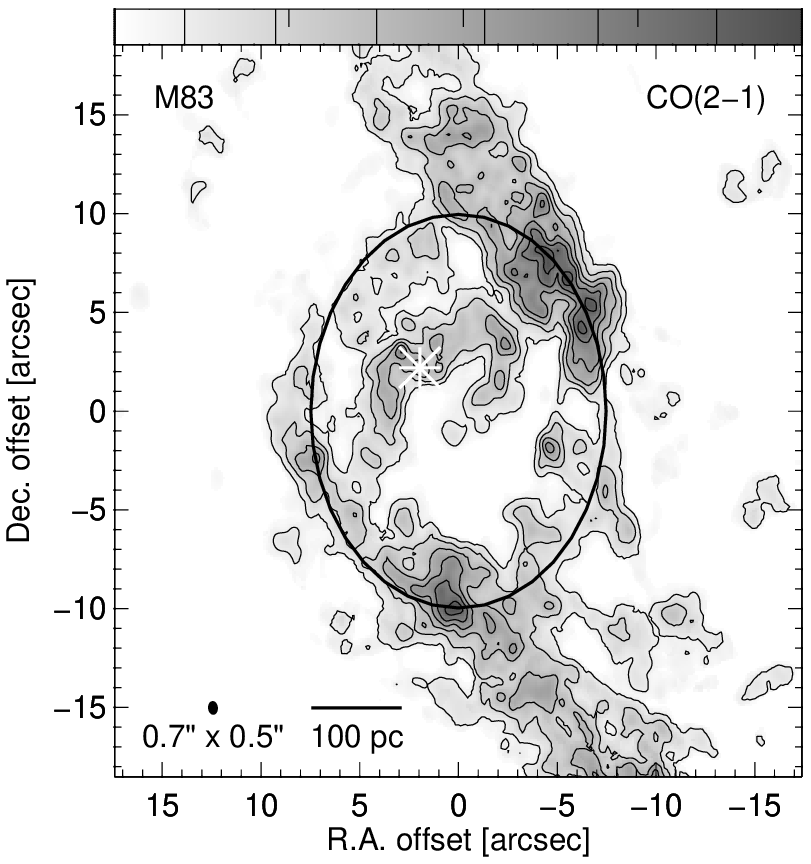}  
\plotone[height=6cm]{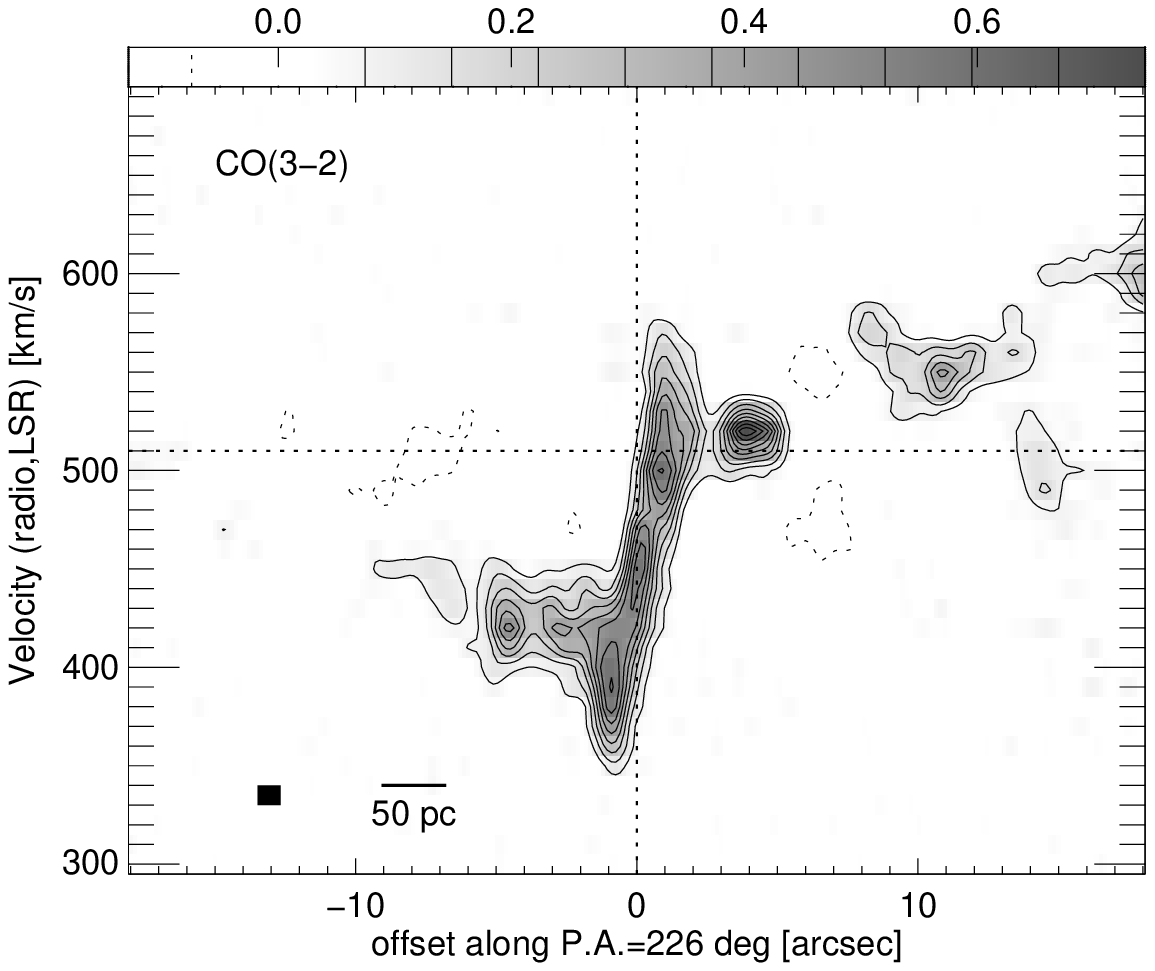}  
\caption{Molecular gas in the center of M83. 
(left) CO(2--1) map showing the circumnuclear molecular ring outlined with an oval.
The galaxy nucleus, or the brightest peak in near infrared, is at the position of the white asterisk. 
It is offset from the center of the molecular ring. 
(right) CO(3--2) position-velocity diagram across the near-IR nucleus.
The steep velocity gradient across the nucleus indicates that the offset nucleus 
is indeed a significant mass concentration and presumably a genuine galaxy nucleus,
although displaced from the geometrical center of the galaxy.
\label{fig.M83}}
\end{figure}

Fig. \ref{fig.M83} shows molecular gas data obtained with the SMA in the center of the nearby spiral galaxy M83.
The barred spiral galaxy has a circumnuclear ring of molecular gas as seen in the left panel.
Such a ring is often seen in barred galaxies.
What is unusual in M83 is its nucleus, or the brightest peak in the near infrared light, 
shown with an asterisk in the CO map.
It is not at the center of the circumnuclear ring but is offset from it by some 70 pc on the sky.
The nucleus is also offset from the geometrical center of the K band isophotes in the kpc scale area around the
central region \citep{Thatte00,Knapen10}.
The molecular ring and the K band light together suggest that the galaxy's center of gravity at the scale of
the molecular ring and beyond is at the center of the ring and not at the visible nucleus.
The offset nucleus, however, has molecular gas rotating around it. 
The CO position-velocity diagram in the right panel of Fig. \ref{fig.M83} shows steep velocity gradient
across the offset nucleus and also shows the absence of such gradient at the center of the molecular ring.
The molecular gas rotating around the offset nucleus and the overall blueshift of the nucleus with 
respect to the systemic velocity of the galaxy were already inferred in lower resolution
SMA data \citep{Sakamoto04} and are confirmed with these new observations.
The cause of this offset nucleus is still under debate and may involve a minor merger 
(\citealt{Rodrigues09}, but see also \citealt{Houghton08}).
It is interesting to note in this context that the size scale of the mini gas disk around the displaced nucleus
($\sim$50 pc) and the dynamical mass of the nucleus (a few times $10^{8} M_{\odot}$ in the central 40 pc)
are both on the same order of the corresponding parameters for the western nucleus of Arp 220.

%%%%%

%%%%%%%%%%%%%%%%%%%%%%%%%%%%%%%%%%%%%%%%%%%%%%%%%%%%
\section{Concluding Remarks}
I have shown examples of high-resolution (sub)millimeter observations of mergers and possible mergers 
and illustrated that such observations can uncover the effect of galaxy merging 
to the dynamics of the interstellar medium and the energy-generating activities in the system. 
The latter effect is caused 
by the fueling to the activities through gas concentration 
and also
by quenching of the activities through gas dispersal via outflows.
There is no doubt that similar observations of mergers are going to lead us to deeper understanding
of galaxy evolution thorough mergers. 
The arrival of ALMA is particularly encouraging for the future of such observational studies.

%%%%%%%%%%
\acknowledgements 
The Atacama Large Millimeter/submillimeter Array (ALMA), an international astronomy facility, 
is a partnership of Europe, North America and East Asia in cooperation with the Republic of Chile. 
This paper makes use of the following ALMA Science Verification data: ADS/JAO.ALMA\#2011.0.00002.SV

%%%%% references

\bibliography{sakamoto}

\end{document}